# Partition laser assembling technique

Yueqiang Zhu [1], Chen Zhang[1]*, Ce Zhang[1], Lijing Zhong[2], Baiqiang Yang[1], Jianrong Qiu[3], Kaige Wang[1]*, Jintao Bai[1], Wei Zhao[1]*

[1]State Key Laboratory of Photon-Technology in Western China Energy, International Collaborative Center on Photoelectric Technology and Nano Functional Materials, Key Laboratory of Optoelectronic Technology of Shaanxi Province, Institute of Photonics & Photon Technology, Northwest University, Xi'an 710127, China

[2]Institute of Light+X Science and Technology, Faculty of Electrical Engineering and Computer Science, Ningbo University, Ningbo 315211, China

[3]State Key Laboratory of Extreme Photonics and Instrumentation, College of Optical Science and Engineering, Zhejiang University, Hangzhou, 310027, China

* Corresponding author. Email: nwuzchen@nwu.edu.cn; wangkg@nwu.edu.cn; zwbayern@nwu.edu.cn

**Abstract:** The advancement of micro/nanofabrication techniques with high throughput, efficiency, and flexibility is critical for fields like integrated photonics, biosensing, and medical diagnostics. This study presents Partition Laser Assembling (PLA), a novel laser technique for fabricating complex micro/nanostructures akin to puzzle pieces. By dividing the target patterns described by scalable vector graphics into partitions, any structures in each partition can be fabricated via structured lights as "light stamp" through spatial light modulation. Unlike traditional direct laser writing, PLA eliminates reliance on mechanical components, avoiding step-like artifacts and ensuring smoother fabrication of complex micro/nanostructures. By seamlessly assembling basic shapes, PLA achieves intricate structures like micro artworks and metalenses with unmatched precision and resolution. Leveraging two-photon fabrication, PLA guarantees high resolution and structural integrity, positioning it as a transformative tool for nanoscale 3D printing. With applications spanning research and industry, PLA paves the way for advanced optical devices, micro/nanofabrications, and next-gen manufacturing technologies.

Laser micro/nanofabrication technology has become a corner stone of future science and industry[1-4], in broad areas, for instance, photonics chip, advanced sensing and biomedical devices. It has been a long-term interest to reach complex structure fabrication with high efficiency, performance and flexibility.

Many fabrication approaches have been developed in the past several decades, including mask-based diffraction lithography[5], electrospinning[6-8], holographic lithography[9], template-based deposition [10], self-rolled-up driven by a stressed nanomembrane[11,12], nanoimprint[13] and direct laser writing (DLW) [14]. Femtosecond DLW is a promising approach with great advantages such as high resolution, true-three-dimensional (3D) processing ability, and processing flexibility [15-17].

Nevertheless, the single-point scanning scheme of femtosecond DLW presents both advantages and limitations. On one hand, it provides exceptional compatibility and flexibility for fabricating complex structures. On the other hand, it inherently limits fabrication efficiency due to its sequential nature[18,19]. More critically, the point-to-point writing process can always produce serrated or stair-stepping flaws, which are fatal in photonic devices, e.g. optical waveguide. Although leveraging femtosecond laser two-photon micro/nanofabrication through multiple beams by light field modulation technology [20,21] enables high speed and parallelly direct writing of micro/nanostructures [22,23], they still struggle with the step-like artifacts and discontinuities between structures inherent to traditional point-by-point fabrication. These limitations hinder the creation of complex cross-scale gradient structures with smoothly curved nanoscale features.

In this study, we present an innovative approach to the next - generation laser fabrication technique, namely the Partition Laser Assembling (PLA) technique, specifically designed for the rapid, flexible, and high - performance production of arbitrary and complex micro/nanostructures (see Movie S1 for reference). It allows for the fabrication of complex geometries that cover multiple scales and morphologies (Fig. 1a), including curved and irregular shapes. During the structural design process, we utilize Scalable Vector Graphics (SVG) (Fig. 1b). In this process, intricate structures are decomposed into fundamental geometric features, such as line segments, circles, squares, and cubic Bézier curves (Fig. 1c), which serve as the basic building blocks for constructing the final design.

The beams corresponding to these basic shapes can be generated using a liquid crystal phase - only SLM. The CGH is rapidly computed through non - iterative methods[24] (Fig. 1d). For instance, when an arbitrary curved structure is required, a corresponding curved beam (e.g., Fig. 1c, 4th row) can be immediately generated by uploading the CGH into the SLM. Since the curved beams have smooth shapes (Fig. 1e and f), serrated or stair - stepping flaws can be avoided during the fabrication process. By assembling all the partitions of basic shapes, a complex micro/nanostructure with a smooth morphology can be ultimately realized.

Fig. 1g depicts the optical system for the PLA technology (for more detailed information, refer to Supplementary Material 1). The fabrication task is organized using SVG as well. Different from the traditional image format that depicts light intensity distribution, SVG additionally provides the shape and path information of each basic structure, with the feature of scale invariance. This enables us to easily determine what kind of light beam should be used and when it should be applied. Therefore, a micro/nanostructure can be fabricated using SVG and light field modulation without the need for laborious scanning processes.

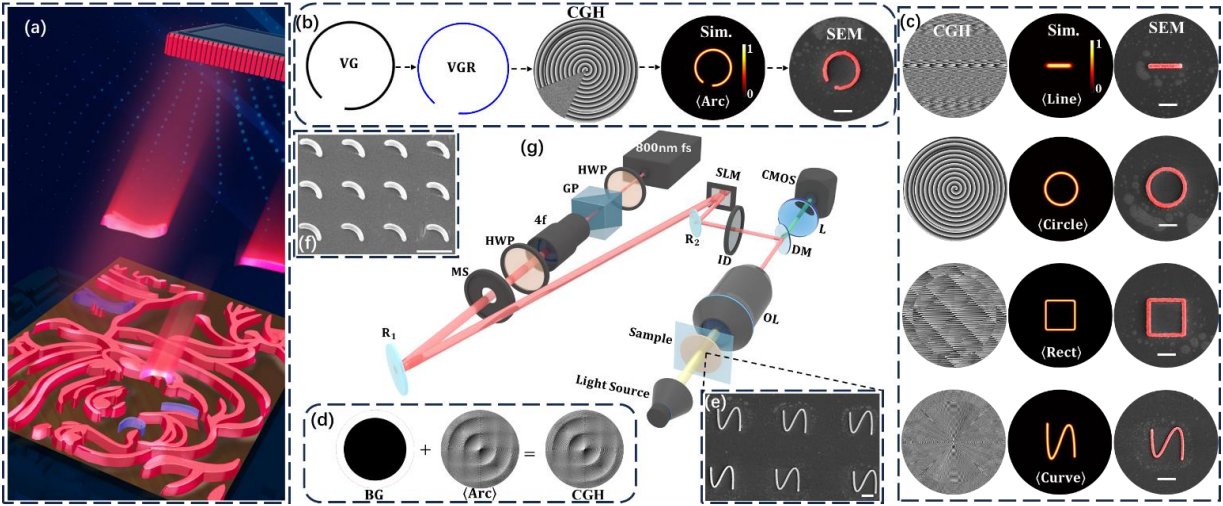

Fig. 1. Rapid fabrication of basic structures of SVG based on PLA technology. (a) is schematic diagram of the principle of PLA technology. (b) fabrication of an arc SVG structure with PLA technology. (c) shows the computer-generated hologram (CGH) of the label structure, the simulated light intensity distribution and the fabrication result, respectively. The microstructures are fabricated in 17 ms. (d) Illustration of the CGH designed by combining the phase of Bessel-Gaussian (BG) beam and label phase. (e, f) Experimental results. (g) is experimental setup. HWP, half-wave plate; GP, Glan prism; 4f, 4f system; MS, mechanical shutter; $R_1$ and $R_2$, reflective mirrors; SLM, liquid-crystal-on-silicon spatial light modulator; ID, iris diaphragm; DM, dielectric mirror; L, lens; CMOS, CMOS camera; Here, microstructure is fabricated with high-NA oil-immersion objective lens (Olympus APO 100X NA 1.25). The scale bar is 3 μm.

Since the light field modulation range is limited, typically 10 μm×10 μm in the *x* - *y* plane for a NA 1.25 objective lens, we need to divide the processing methods for the fabricated structure into two situations. (1) For targets smaller than the movement range of the light field modulation, partitioning is not executed; (2) For targets larger than the movement range of the light field modulation, partitioning is carried out.

As an example, we first demonstrate how to fabricate a smaller target structure using the PLA technique and present the preliminary results. The workflow of the PLA technique is shown in Fig. S4, where a small-scale fingerprint structure is fabricated and presented. As is well-known, no two people have the same fingerprints because of the tiny yet distinct differences in the curves of their fingerprints. Here, as an example, we show that a replica of a fingerprint can be simply and accurately fabricated using the PLA technique, for potential application in bioinformation storage.

Before fabrication, we interpret the SVG of the fingerprint (Fig. 2a) into mathematical equations and construct the outline of the fingerprint (Fig. 2b). Subsequently, the outline of the fingerprint is filled to reconstruct the structure (Fig. 2c). The reconstructed fingerprint is converted into path label structures to obtain cubic Bézier curves (Fig. 2d). Thereafter, a series of CGHs (Fig. 2e) are generated according to the path label structures and non-iterative algorithms (see the supplementary material) to modulate the laser beams as required (Fig. 2f). After fabricating all the corresponding structures, the fingerprint is assembled (Fig. 2g).

In this process, only 20 frames of CGH are used. The exposure time for a single exposure is 17 ms, and the total fabrication time is only 0.3 s. The fabricated fingerprint structure is 7 μm wide, with a minimum linewidth of 300 nm. The fabricated structure exhibits good surface flatness, uniformity, and repeatability (Fig. 2h), with negligible assembling deviation.

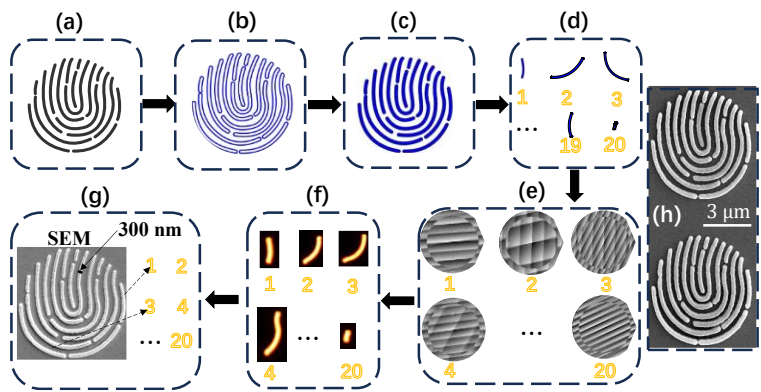

**Fig. 2.** Fabricating a fingerprint utilizing PLA technique, (a) is SVG images, (b) represents the contour line of the reconstructed SVG, (c) denotes the reconstructed SVG obtained after processing the filled area, (d) interpreting of the reconstructed SVG into path label structures to get cubic Bézier curves, (e) are CGH obtained based on path label structures and non-iterative algorithms method, which 17 frames CGHs are obtained, (f) are light intensity distribution generated by corresponding CGH in (e), (g, h) is the final fabrication results by loading 20 frames CGH into SLM. The exposure time for a single CGH is 17 ms.

For large-area fabrication, a partition process is employed to divide the large structure into smaller partitions. This process is influenced by several factors, including the specifications of the optical system (e.g., laser power, wavelength, and objective lens) and the characteristics of the target structures to be fabricated (e.g., resolution and structure size).

To demonstrate the capability of the PLA technique for large-area, complex, and non-uniform structure fabrication, we selected a segment extracted from the renowned artwork "Along the River during the Qingming Festival" (Fig. 3a). This example aims to illustrate that any structure designed using SVG can be automatically fabricated via the PLA technique, thereby showcasing its readiness for industrial applications, from initial modeling to final fabrication. Moreover, the technique preserves the original structures with nano-resolution.

As shown in Fig. 3(a-i), a portrait of an old fisherman (Fig. 3a), which is a small part of the artwork, is selected for fabrication. Given that the structure (Fig. 3b) has dimensions of 30 μm × 25 μm, which exceed the modulation area of 10 μm × 10 μm, the portrait is divided into nine partitions (Fig. 3c). Since the structures are continuously distributed, partitioning may disrupt their continuity, potentially lead to assembly deviations, which pose a significant challenge to fabrication quality. Additionally, the varying line widths of the structures present another challenge for the PLA technique.

The fabrication sequence is schematically illustrated in Fig. 3(d). In each partition, the structures are decomposed into basic shapes (Fig. 3e), which are then assembled into local structures within the partitions (Fig. 3f). Subsequently, CGHs (Fig. 3g) are computed using a non-iterative method and uploaded to the SLM to generate the corresponding structured light.

The fabrication result is presented in Fig. 3i. The fabricated structures exhibit high consistency with the original portrait (Fig. 3a), indicating a high level of fabrication quality. No significant assembly deviation was observed. Ultimately, a large-scale structure with a typical width of 143 nm was successfully implemented.

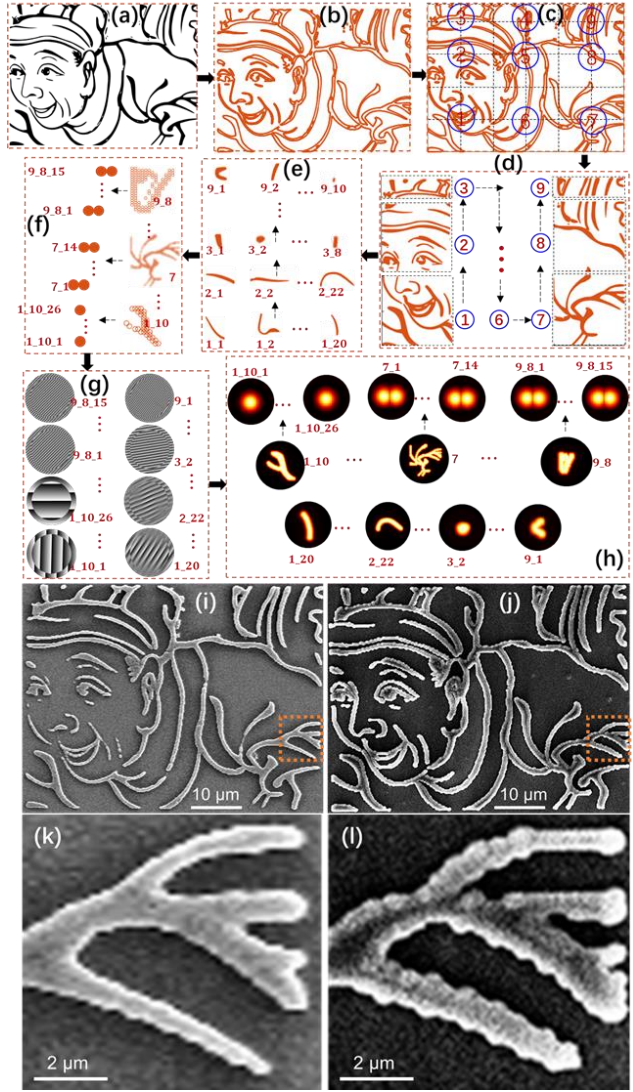

**Fig. 3.** Fast processing of complex microstructure by the PLA technology, (a-h) Process of fabricating complex structure requires 345 frames of CGH, with an exposure time of 17 ms per frame of CGH based on PLA technology. (i, j) Experimental results obtained by PLA and LDW respectively. (k, l) are the local views of (i, j) respectively.

In contrast to the progressive scanning method used in conventional laser direct writing (LDW) (Fig. 3j and l), the PLA technique offers higher fabrication efficiency and superior quality. All the fabricated curved structures exhibit smooth variations (Fig. 3k), without the serrated or stair-stepping flaws fabricated by conventional LDW (Fig. 3l). This smoothness is crucial for the production of photonic chips, where complex structures with varying linewidths are often required. Relying on the two-photon absorption of the material, the PLA technique can achieve fine structures with linewidths as small as ~75 nm (Fig. S6). Technically, the linewidth could be further reduced if higher-order nonlinear absorption or super-resolution methods is applied.

To demonstrate the potential application of the PLA technique in micro/nanofabrication, a representative geometric meta-lens for beam focusing was designed and fabricated in this study. The meta-lens consists of elliptical nanorods (Fig. 4a) arranged periodically. The meta-lens is designed to focus a 639 nm circularly polarized beam with a focal length of $f = 60$ μm (Fig. 4b).

The phase distribution profile of the nanorods, generated using Eq. S1, is shown in Fig. 4c, with a phase range from 0 to π. The distribution of the nanorod matrix was then designed using Finite-Difference Time-Domain (FDTD) analysis.

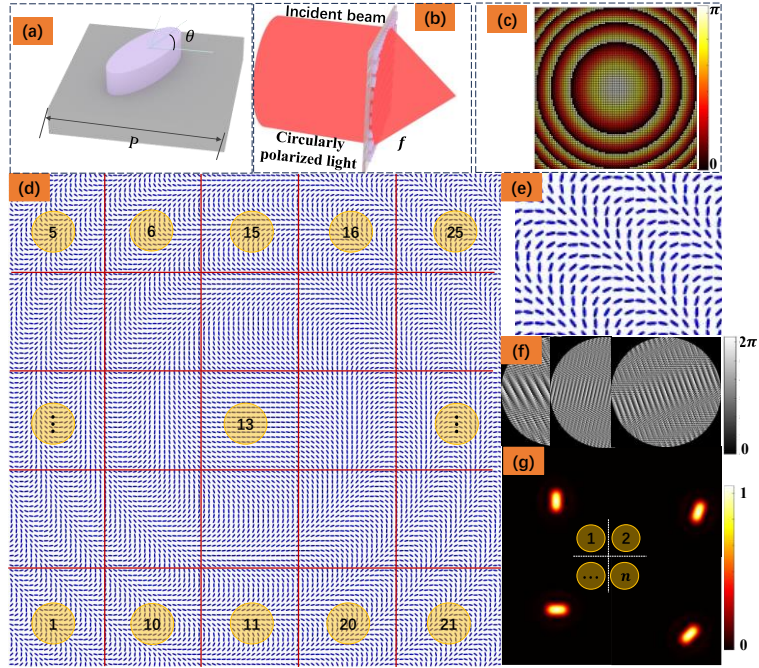

Fig. 4 A meta-lens for focusing designed according to Pancharatnam-Berry [25] phase. (a) Structure of the menta-lens unit; (b) Schematic illustrating the focusing mechanism of the menta-lens, where a circularly polarization beam is directed onto the menta-lens and converges at the focal point; (c) Orientation profile distribution of the nanorods; (d) Nanorod distribution was first obtained by FDTD simulations. Then the vector pattern distribution was reconstructed using PLA technology. The distribution divided into 25 partitions; (e) Enlarged partition 25 of (d); (f) A series of phase distributions derived from the PLA technology; (g) Corresponding light intensity distributions of the nanorods obtained from the phase distributions.

The designed sketch of the nanorod matrix structures was further converted into SVG, which contains position and angular information. Given that the size of the meta-lens exceeds the modulation range of the optical system, the meta-lens sketch was divided into 25 partitions, as shown in Fig. 4d. The fabrication process followed a zigzag sequence: partitions 1 to 5 were fabricated first, followed by partitions 6 to 10, and so on until all the partitions were completed. An enlarged view of partition 25 is shown in Fig. 4e. Within each partition, the CGHs of the nanorods were generated (Fig. 4f). By loading the phase maps into the SLM, the light intensity distribution of each nanorod was obtained, as shown in Fig. 4g. A total of 10,000 frames of CGHs were generated during this process. Finally, with the laser power set to 14.3 mW, dynamic fabrication was performed in each subregion until all structures were completed.

The meta-lens has a diameter of 43 μm (Fig. 5a). The nanorod structures have dimensions of 459 nm ± 54 nm in length and 197 ± 36 nm in width. The average distances between nanorods are 592 nm and 575 nm in horizontal and vertical directions respectively. The minimum linewidth and gap between adjacent nanorods are 134 nm and 74 nm (Fig. 5b), respectively, showing the high fabrication accuracy of PLA technique.

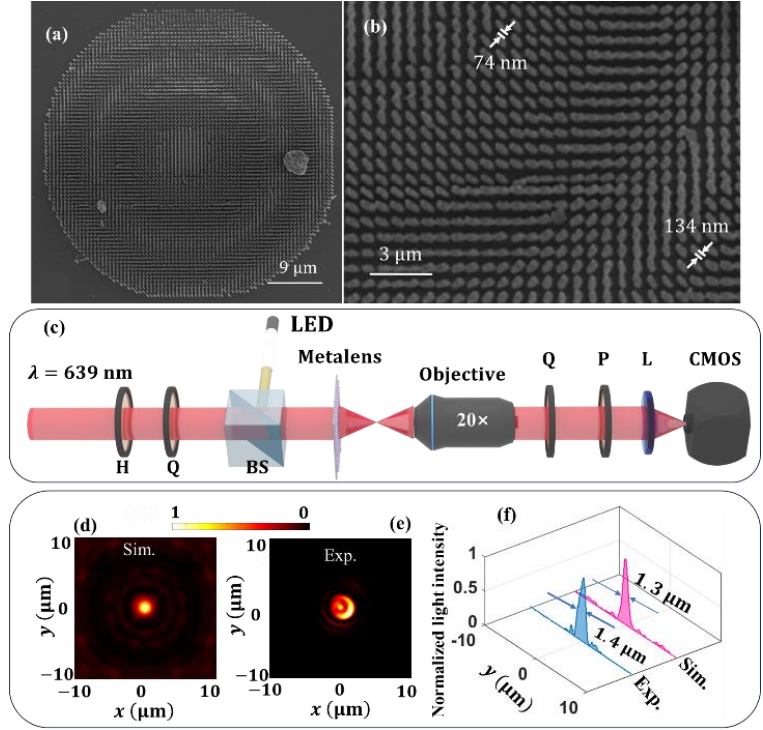

Fig. 5 Meta-lens fabricated by PLA (a) and its zoom-in view (b) scanned by SEM. The exposure time for a single CGH is 17 ms. (c) Optical setup for meta-lens test. Focal spots between simulation (d) and experiment (e). (f) Light intensity profiles in simulation (d) and experiment (e).

The performance of the meta-lens was evaluated using the experimental setup depicted in Fig. 5c. A collimated continuous-wave laser with a wavelength of 639 nm was first passed through a half-wave plate and a quarter-wave plate to generate circularly polarized light. This circularly polarized beam was then directed into a beam splitter and subsequently illuminated the meta-lens. To aid in locating the meta-lens structure, an LED white light source was also employed. The laser beam passing through the meta-lens was collected by an objective lens to form a 4f optical system. After collimation, the beam passed through a quarter-wave plate and a polarizer before being focused onto a CMOS camera for beam analysis.

The focusing performance of the meta-lens, characterized by the light intensity distribution, is shown in Fig. 5e and f. The full width at half maximum (FWHM) of the focused spot, as measured from the cross-sectional profile of the point spread function along the $y$-axis, is 1.4 µm, which is in good agreement with the simulation results. While the PLA technique enabled precise control over the orientation and position of each unit cell, some cross-linking between nanorods was observed. This cross-linking may be the primary cause of the deviation between the experimental results and the simulations.

The PLA technique is currently in its nascent stage, and the fabrication time remains relatively long due to two primary factors. One is the time required for analyzing the fabrication task (encompasses SVG decomposition, partitioning, and parametric analysis). Particularly, the automatic adjustment of laser power according to the basic shape poses a considerable challenge, which could be resolved by artificial intelligence in future. The other is the slow refresh rate of a commercial SLM at 60 Hz, which is a major bottleneck in improving the fabrication efficiency of the PLA technique. We hope that the development of high-speed SLMs will significantly alleviate this limitation and enhance the overall efficiency of the PLA process.

In this context, we report a novel laser fabrication method termed the PLA technique, which integrates two-photon absorption, light field modulation, and a vector path guided by SVG for rapid and flexible nanofabrication. As a preliminary test of this method, we fabricated a complex structure using various basic shape structures and optimized light field modulation to improve fabrication quality and time efficiency. A minimum linewidth of 75 nm was achieved.

Despite the imperfections of the current PLA technique, it represents an effective approach for the development of the next generation of laser fabrication systems. The method highlights the potential for advancing PLA techniques toward fully automated, all-electric systems. This advancement would provide a fast and efficient lithography technology capable of engraving complex structures with high precision, thereby paving the way for future advancements in nanofabrication.

**Acknowledgements:** This investigation is supported by National Natural Science Foundation of China (Grant No. 51927804, 61775181, 61378083).

**Author contributions.**

Conceptualization: Wei Zhao

Methodology: Yueqiang Zhu, Wei Zhao

Investigation: Yueqiang Zhu, Chen Zhang

Visualization: Yueqiang Zhu

Funding acquisition: Chen Zhang, Kaige Wang, Jintao Bai, Wei Zhao

Project administration: Kaige Wang, Jintao Bai, Wei Zhao

Supervision: Kaige Wang, Wei Zhao

Writing – original draft: Yueqiang Zhu, Chen Zhang, Ce Zhang, Lijing Zhong, Baiqiang Yang, Jianrong Qiu, Kaige Wang, Jintao Bai, Wei Zhao

Writing – review & editing: Yueqiang Zhu, Chen Zhang, Ce Zhang, Lijing Zhong, Baiqiang Yang, Jianrong Qiu, Kaige Wang, Jintao Bai, Wei Zhao

**Competing interests:** The authors declare no conflicts of interest.

**Data and materials availability:** The data presented in this study are available on request from the corresponding author.

**Supplementary Materials:**
    Principles, methods and materials
    Supplementary texts and data

Movie S1